\documentclass[aps,prl,twocolumn,showpacs,amsmath,amssymb]{revtex4}
\usepackage{graphicx}
\usepackage{dcolumn}
\usepackage{bm}
\begin{document}

\title{
Drag force in bimodal  cubic-quintic non-linear Schr\"odinger
equation
}
\author{ David Feijoo, Ismael Ord\'o\~nez, Angel Paredes, 
Humberto Michinel}
\affiliation{ \'Area de \'Optica, Departamento de F\'\i sica Aplicada,
Universidade de Vigo, As Lagoas s/n, Ourense, ES-32004 Spain;}

\begin{abstract}
We consider a system of two
 cubic-quintic non-linear Schr\"odinger
equations in two dimensions, coupled by repulsive cubic terms. 
We analyse situations in which a probe lump of one of the modes
is surrounded by a fluid of the other one and analyse their interaction. 
We find a realization of D'Alembert's paradox for small velocities and non-trivial drag forces for
larger ones.
We present numerical analysis including the search of static and traveling form-preserving solutions 
along with simulations of the dynamical evolution in some representative examples.

\end{abstract}


\maketitle


\section{I. Introduction}

The nonlinear Schr\"odinger equation (NLSE) is a widely used model in the study of 
quasi-monochromatic, 
nonlinear dispersive waves. Among other applications, it describes the dynamics of
Bose-Einstein condensates (BECs) in the mean field approximation \cite{dalfovo}
or the propagation of laser beams in optical fibers \cite{agrawal},
for which different expressions have been used for the nonlinearity of the refractive index \cite{review}. 
The cubic-quintic NLSE \cite{cq-old}
is arguably the simplest model for competing nonlinearities \cite{compete} and has been used in
many different contexts, see {\it e.g.} \cite{CQ,Josserand,critical,rarefaction} and references therein.
In two transverse dimensions, which is a relevant case for non-linear optics,
 the cubic (focusing)-quintic (defocusing) model (CQNLSE)
 presents remarkable features.
There are families of stable solitary waves (solitons and vortices) which become flattop when the power of the
beam is large: the propagation constant  never exceeds a critical value $\beta_{cr}$ and, for growing
power $\beta\to {\beta_{cr}}^-$, there is a growing region where the amplitude also tends to a critical value
 $\psi \approx e^{i\beta_{cr}z} \psi_{cr}$, as  was established by different
 numerical and analytical methods in \cite{crit1}
  and later rigorously proved  in \cite{critical}. This behaviour endows the flattop solutions of the
  CQNLSE with the properties of a liquid \cite{liquid}. The $|\psi| \approx \psi_{cr}$ is a region of
  constant pressure and the rapid decay from $|\psi| \approx \psi_{cr}$ to $|\psi| \approx 0$ can be identified
  with a liquid-vapour interface characterized by a surface tension, leading to effects analogous
  to capillarity and dripping in regular liquids \cite{dripping}.
Remarkably, the first neat experimental realization of this liquid of light has been reported recently \cite{experiment},
  following the proposal of \cite{coherent-media} of engineering the desired optical properties in a coherent medium.
    
A natural question is whether there are other hydrodynamical properties that can be defined for this kind of solutions
of the CQNLSE. 
In this paper we analyse the drag, namely the force which opposes to the motion of an object within a surrounding
fluid. An ``object'' inside the fluid described by a NLSE can be modelled by implementing appropriate boundary conditions
at the edge of the moving body, as was done in the framework of superfluidity in \cite{Josserand,drag}, where
similar questions to those addressed here were studied. 
We will consider a different approach which might be  suitable for nonlinear optics or BECs:
the probe object is also described by a CQNLSE, leading to a bimodal system of coupled equations for two
wave-functions $\psi_1$, $\psi_2$. 
In non-linear optics \cite{manakov}, the $\psi_i$ typically correspond to different polarizations or
different carrier wavelengths  while in
BECs they represent different atomic species in the condensate \cite{bec1} or different internal
states of the same isotope \cite{bec2}, see {\it e.g.} \cite{bec3} and references therein.

The system of equations we will study is the following:
\begin{eqnarray}
-i\partial_z \psi_1=\nabla^2 \psi_1 + (|\psi_1|^2 - |\psi_1|^4 - \gamma\, |\psi_2|^2)\psi_1  \nonumber \\
-i\partial_z \psi_2=\nabla^2 \psi_2 + (|\psi_2|^2 - |\psi_2|^4 - \gamma\,|\psi_1|^2)\psi_2 
\label{eqs}
\end{eqnarray}
where for simplicity we have fixed to unity several coefficients. 
The laplacian is taken over two transverse dimensions $\nabla^2 = \partial_x^2 + \partial_y^2$. 
For the crossed interaction,
we only introduce cubic terms weighed by a constant $\gamma$. We will restrict ourselves to
analysing $\gamma>0$, namely
inter-modal repulsion resulting in a fluid with inmiscible phases, which is the most suitable situation
to formulate thought experiments regarding drag forces.

A bimodal cubic-quintic model similar to (\ref{eqs}) was first introduced in
\cite{bimodal99} to discuss the interaction between solitons of both species.
Variations of this model were later used for the study of vector solitons \cite{bimodalcq},
their dynamics \cite{wangtian} 
and modulational instability \cite{modul1,bimodal-discrete}. It is worth pointing out that 
these works mostly deal with intermodal attraction $\gamma<0$. An exception is \cite{bimodal-discrete}, which deals
with BECs where interspecies forces can
be tuned using Feshbach resonances and can be either
attractive or repulsive.

In Eqs. (\ref{eqs}),
the norm $\int |\psi_i|^2dx dy$ for each
species is conserved separately upon evolution in $z$. 
Moreover, it is straightforward to check that total momentum is preserved:
\begin{eqnarray}
\vec p=\frac{1}{2i}\int \sum_{i=1,2} \left(\psi_i^* \vec \nabla \psi_i - \psi_i \vec \nabla \psi_i^* \right) dx\,dy
\end{eqnarray}
but the $\vec p_i$ associated to each species are not separately conserved, {\it i.e.},
momentum can be transferred between species, leading to inter-modal macroscopic forces. 
In the following, we will consider the dynamics of a droplet of $\psi_1$ surrounded by a large background of $\psi_2$,
with
$\int |\psi_2|^2 dS \gg \int |\psi_1|^2 dS$. We thus study the effects of the drag force exerted
by the $\psi_2$-fluid on an $\psi_1$-probe ``object''. 

In section II, we discuss the static solutions. In section III, we find form-preserving traveling solutions
which can be interpreted as dragless motion of an object within an inviscid fluid and are therefore
related to D'Alembert's paradox. These configurations exist below some limiting velocity.
We also discuss how this kind of solutions can be approached in processes
with dynamical
evolution.
In section IV, we devise a kind of thought falling ball viscometer experiment and introduce an approximate analogy between this 
intricate nonlinear setup  and a simple mechanical system. 
In section V, we consider a case in which both species are initially separated and show the similarity of simulated processes
with the entrance of a rigid object in a liquid. Finally, we present our conclusions in section VI.

\section{II. Static solutions}

We start by looking for radially symmetric,
stationary solutions with a circle of $\psi_1$ surrounded by an infinite critical background of 
$\psi_2$, with $\int |\psi_2|^2 dS= \infty$. Namely:
\begin{eqnarray}
\psi_1 = e^{i\,\beta_1\,z} f_1(r)\,,\qquad\psi_2=e^{i\,\beta_{cr}\,z} f_2(r)
\end{eqnarray}
where $f_1(r)$, $f_2(r)$ are real functions and 
 $\beta_{cr}=\frac{3}{16}$ \cite{critical}.
 The
system (\ref{eqs}) is reduced to:
\begin{eqnarray}
\partial_r^2 f_1 + r^{-1}\partial_r f_1=\beta_1 f_1 - (f_1^2 - f_1^4 -\gamma f_2^2)f_1\,\,,\nonumber\\
\partial_r^2 f_2 + r^{-1}\partial_r f_2=\beta_{cr} f_2 - (f_2^2 - f_2^4 -\gamma f_1^2)f_2\,\,.
\label{eqsf}
\end{eqnarray}
 Boundary conditions at infinity are:
\begin{equation}
\lim_{r\to\infty}f_1(r)=0\,,\qquad\lim_{r\to\infty}f_2(r)=\psi_{cr}=\sqrt3/2
\label{bc}
\end{equation}
The profile of the functions at $r\to\infty$ consistent with (\ref{bc}) can be found by computing the
leading terms in (\ref{eqsf}). We find that $f_1(r)\sim r^{-1/2}\exp(-\sqrt{\beta_1+3\gamma/4}\,r)$. 
Therefore, solutions can only exist for $-\frac34\gamma <\beta_1$.
The function 
$f_2$ behaves as $\sqrt3/2 - f_2(r)\sim r^{-1/2}\exp(-\sqrt{3}r/2)$ if $\beta_1 >\frac{3}{16}(1-4\gamma)$ and as
$\sqrt3/2 - f_2(r)\sim \exp(-2\sqrt{\beta_1+3\gamma/4}\,r)$ for $\beta_1 \leq \frac{3}{16}(1-4\gamma)$.

At $r=0$, solutions must be regular $f_1'(0)=f_2'(0)=0$. By performing a Taylor expansion, we find that
near the origin the functions can be written in terms of two constants:
\begin{eqnarray}
f_1(r)=a_0 + \frac{a_0}{4}(\beta_1 -a_0^2 + a_0^4 + \gamma\, b_0^2)r^2+{\cal O}(r^4) \nonumber\\
f_2(r)=b_0 + \frac{b_0}{4}(\frac{3}{16} -b_0^2 + b_0^4 + \gamma \, a_0^2 )r^2+{\cal O}(r^4) \nonumber
\end{eqnarray}

For a given $\gamma$, numerical solutions
of (\ref{eqsf}), (\ref{bc})
 can be found by rewriting the equations in a finite differences scheme and solving the resulting
non-linear algebraic system by standard methods. 
There is a one-parameter family of nodeless monotonic solutions
($f_1'(r)<0$ and $f_2'(r)>0$ $\forall r>0$) with $-\frac34\gamma <\beta_1<\frac{3}{16}$.
In the limit $\beta_1 \to \frac{3}{16}$, the norm of $\psi_1$ diverges and one has a kink-antikink solution
with two separate ``liquids''. Naming $r_*$ the radius of the region where 
$f_1(r)$ dominates, we have $r_* \to \infty$ as $\beta_1 \to \frac{3}{16}$, $f_1(r)\approx \psi_{cr}$ for $r \ll r_*$ and $f_1(r) \approx 0$
for $r\gg r_*$ and viceversa for $f_2(r)$.

\begin{figure}[htb]
\begin{center}
{\includegraphics[width=0.8\columnwidth]{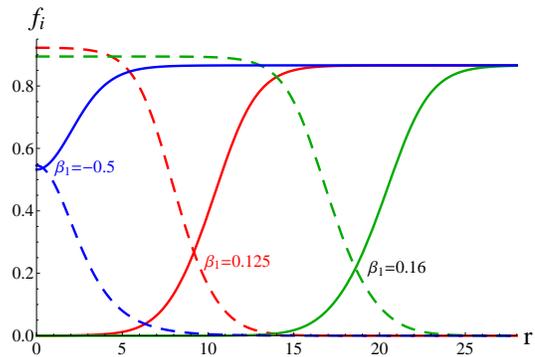}}
\end{center}
\caption{Solutions of (\ref{eqsf}), (\ref{bc}) with $\gamma=1$ for $\beta_1=-0.5$, $\beta_1=0.125$
and $\beta_1=0.16$. For larger $\beta_1$, the value $r_*$ at which $f_1(r)$ decays and $f_2(r)$ rises increases and,
therefore the normalization $\int |\psi_1|^2dS$ also grows.
Dashed lines correspond to $f_1(r)$ and solid lines to $f_2(r)$.
}

\label{fig1}
\end{figure}

\begin{figure}[htb]
\begin{center}
{\includegraphics[width=0.8\columnwidth]{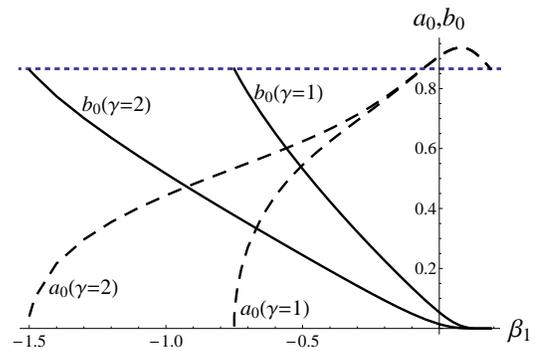}}
\end{center}
\caption{Values of $f_1(0)=a_0$ (dashed lines) and $f_2(0)=b_0$ (solid lines) computed from the numerical solutions for the families
computed taking $\gamma=1$ and $\gamma=2$. The horizontal dotted line marks $\psi_{cr}=\sqrt3/2$.
}
\label{fig2}
\end{figure}

Figure \ref{fig1} shows three examples of $f_1(r)$, $f_2(r)$ pairs computed numerically.
In Fig. \ref{fig2}, we depict the values $f_1(0)=a_0$, $f_2(0)=b_0$ for the families of solutions with $\gamma=1$ and
$\gamma=2$.

In the following sections, we will take $\gamma=1$. The qualitative results hold for more general values
of $\gamma>0$.

\section{III. D'Alembert's paradox}

We now show that there exist solutions in which the lump of the first species moves with constant velocity 
$U$
within the fluid of 
the second species, {\it i.e.} there are situations in which the drag force is exactly zero.
They correspond to steady flows in the moving reference frame.
A similar behaviour involving a different NLSE model was first found in \cite{drag}.
Notice that we abuse of language using the word velocity to refer to derivatives with respect to $z$, which in the nonlinear
optics framework correspond to propagation distance rather than time. In that case, the variations in $z$ are a consequence
of having non-trivial components of the wave-vector apart from $k_z$ and this velocity is, physically, the propagation angle
with respect to the $z$-axis.
We introduce an ansatz of the form: 
\begin{eqnarray}
\psi_1&=&e^{i\,\beta_{U}\,z} \phi_1(x,y-U\,z)\,,\nonumber\\
\psi_2&=&e^{i\,\beta_{cr}\,z} \phi_2(x,y-U\,z)\,,
\end{eqnarray}
This system of equations can be treated
along the lines of \cite{jr1}:
consider $\eta=y-U\,z$ and write the system as a PDE in $x$, $\eta$. One finds the following: 
\begin{eqnarray}
iU\partial_\eta \phi_1=\nabla^2 \phi_1+\left( |\phi_1|^2-|\phi_1|^4-\gamma|\phi_2|^2- \beta_U \right)\phi_1
\nonumber\\
iU\partial_\eta \phi_2=\nabla^2 \phi_2+\left(|\phi_2|^2-|\phi_2|^4-\gamma|\phi_1|^2-\beta_{cr} \right)\phi_2
\label{eqsphi}
\end{eqnarray}
where $\nabla^2$ should now be understood as 
$\partial_x^2 + \partial_\eta^2$. Boundary conditions at infinity ($x^2 + \eta^2 \to \infty$) are
$\phi_1 \to 0$, $\phi_2 \to \sqrt3/2$. We split real and imaginary parts as:
\begin{equation}
\phi_1 = \phi_{1R}+i\,\phi_{1I}\,,\qquad
\phi_2 = \phi_{2R}+i\,\phi_{2I}
\end{equation}
The system (\ref{eqsphi}) is invariant under $x\to -x$ and under $\eta  \to -\eta$ together with
$\phi_{1I} \to - \phi_{1I}$, $\phi_{2I} \to - \phi_{2I}$. Thus, it is enough to compute the functions
for $x>0$, $\eta>0$ and
solutions must be consistent with the following
set of Neumann and Dirichlet boundary conditions at $x=0$ and $\eta=0$:
\begin{eqnarray}
0&=&\partial_x \phi_{1R}|_{x=0}=\partial_x \phi_{2R}|_{x=0}=\partial_x \phi_{1I}|_{x=0}=\partial_x \phi_{2I}|_{x=0}
\nonumber\\
0&=&\partial_\eta \phi_{1R}|_{\eta=0}=\partial_\eta \phi_{2R}|_{\eta=0}\nonumber\\
0&=& \phi_{1I}|_{\eta=0}= \phi_{2I}|_{\eta=0}
\label{bc1}
\end{eqnarray} 
We have found numerical solutions of the problem (\ref{eqsphi}), (\ref{bc1}) by using a finite difference
method: we discretize the $x-\eta$ plane in a lattice of $N_x \times N_\eta$ points and write down the
resulting
(approximately) $4N_x N_\eta$ algebraic nonlinear equations for the same number of real variables. Given a
judicious initial ansatz,
solutions can be found by a standard Newton-Raphson method.
For fixed $\gamma$, there is a two-parameter family of solutions, depending on $U$ and $\beta_U$. Since
the solutions with $U=0$ have been computed in section II, they are a good starting point to search for different
solutions of the family. In particular, we are interested in solutions with different $U$'s but constant
$\int |\phi_1|^2 dx \,d\eta $. The relation of $\beta_U$ with the norm is
non-trivial, but, for fixed $U$, we can vary $\beta_U$ and compute different solutions until we get the one
with the desired normalization. For fixed normalization, there is a maximal value of $|U|$ for which the
solution exists.

\begin{figure}[htb]
\includegraphics[width=0.48\textwidth]{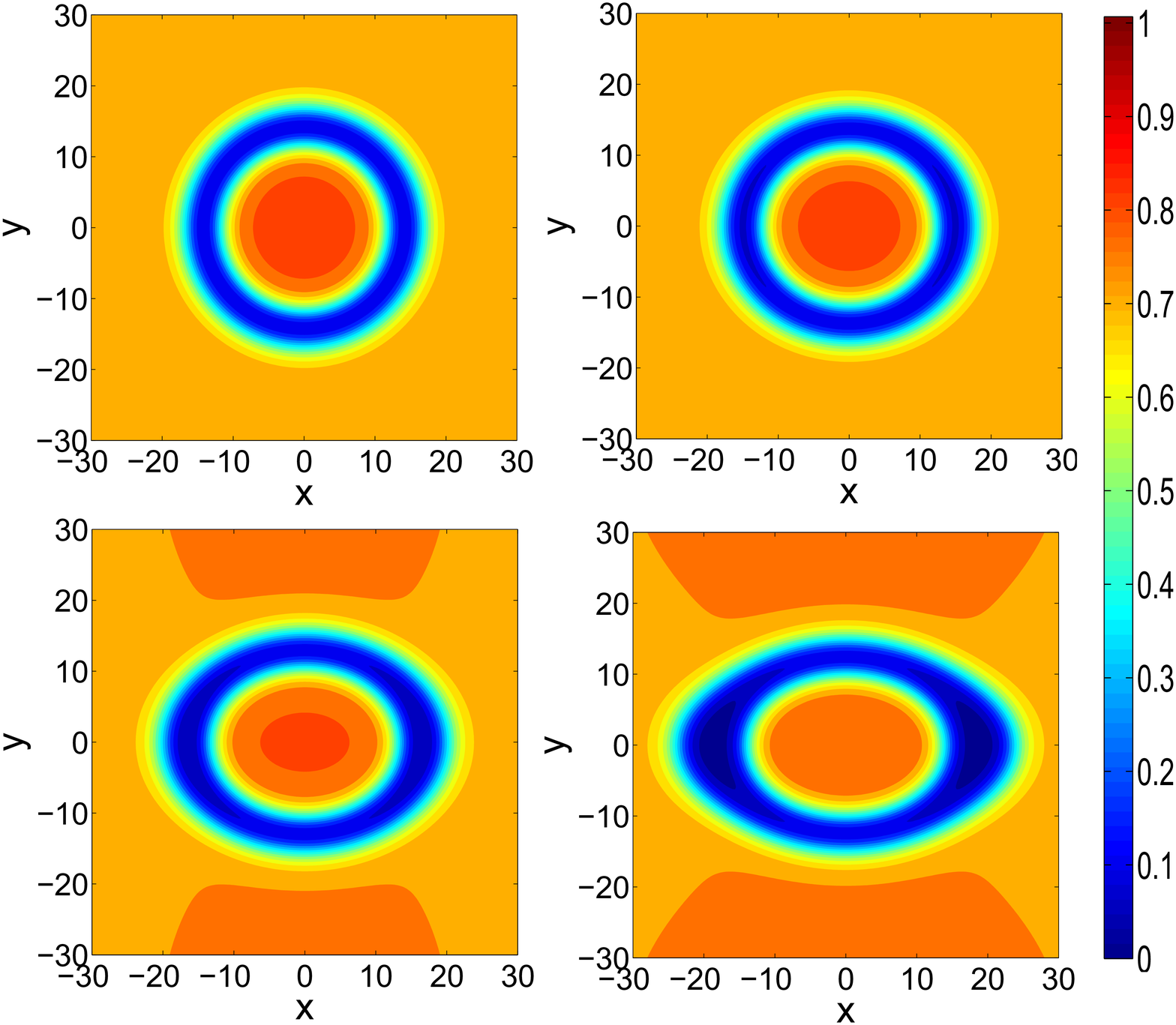}
\caption{Contour plots of $|\phi_1|^2+|\phi_2|^2$ for four solutions of Eqs. (\ref{eqsphi}) with
$\int |\phi_1|^2 dx\,d\eta \approx 355$. This normalization corresponds to $\beta_1=0.15$ in the formalism
of section II.
The four images correspond to $U=0$; $U=-0.1$; $U=-0.15$; $U=-0.17$; respectively. 
}
\label{fig3}
\end{figure}

A few examples of numerical approximations --- computed in a $120\times 240$ lattice --- 
are depicted in Fig. \ref{fig3}.
We depict contour maps of the quantity $|\phi_1|^2+|\phi_2|^2$. It should be understood that the inner 
region mostly corresponds to $|\phi_1|^2$ and the outer one to $|\phi_2|^2$. The region where
$|\phi_1|^2+|\phi_2|^2$ drastically descends is the interface. The plots show how the $|\phi_i|^2$-distributions of the
traveling solutions get deformed as the velocity is increased.

It is also interesting to understand what happens if the initial conditions do not correspond exactly to these
stationary solutions. With that aim, we have performed simulations in which the static solutions of section II
are given a boost, {\it i.e.}, $\psi_1$ is multiplied by $e^{-i\,u_0\,y/2}$ where $u_0$
is (minus) the initial velocity and then used as initial conditions in (\ref{eqs}). The evolution is computed by a standard
split-step pseudo-spectral method, the so-called beam propagation method. In order to avoid spurious 
effects related to boundary conditions, we have taken a finite droplet for the second species.
Simulations \cite{supplemental} show that, initially, the boosted soliton loses momentum to the medium but eventually
tends to a constant velocity, approaching the above described behaviour related to D'Alembert's paradox.

\begin{figure}[htb]
\includegraphics[width=\columnwidth]{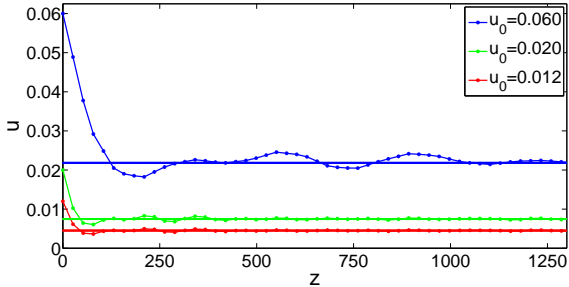}
\caption{Examples of the evolution with $z$ of the velocity of the first species lump immersed in the dragging
fluid. The horizontal lines mark the asymptotic velocity inferred from the simulations.
}
\label{fig4}
\end{figure}

In order to  describe this effect quantitatively,
let us define the central position of the $\psi_1$ droplet as:
\begin{equation}
\langle y_1 \rangle (z) = \frac{\int  \int y |\psi_1(z)|^2 dx dy}{\int  \int |\psi_1(z)|^2 dx dy}
\end{equation}
and its velocity as $u(z)=-\frac{d\langle y_1 \rangle (z)}{dz}$. Figure \ref{fig4} shows how $u$ evolves upon
propagation for different examples.

\section{IV. Terminal velocity and drag force}

We now devise a thought experiment which can be considered an NLSE version of the evolution
of a body moving within a fluid subject to an external force. 
Let us modify (\ref{eqs})
to include an extra term accounting for a potential acting on $\psi_1$ along the $y$-direction --- which in the case of optics would
correspond to a linear variation of the linear refractive index:
\begin{eqnarray}
-i\partial_z \psi_1&=&\nabla^2 \psi_1 + (|\psi_1|^2 - |\psi_1|^4 - \gamma\, |\psi_2|^2)\psi_1 - g \,y\,\psi_1 \nonumber \\
-i\partial_z \psi_2&=&\nabla^2 \psi_2 + (|\psi_2|^2 - |\psi_2|^4 - \gamma\,|\psi_1|^2)\psi_2 
\label{eqs_force}
\end{eqnarray}
We will consider an initially static solution
as discussed in section II,
for which  $g$ is eventually turned on, namely $g=0$ for $z<0$ and is shifted to a constant for $z>0$.
We compute
this evolution by numerically integrating (\ref{eqsf}) by the split-step pseudo-spectral beam propagation method.
The $\psi_1$ distribution starts drifting driven by $g$ but the drag force of the fluid
eventually stops the acceleration
 and the motion tends to a terminal velocity
$u_T(g)$. 

The qualitative behaviour is different for small and large $g$. For large $g$, a void is generated in the wake of
the moving object. For smaller $g$, vortex-antivortex pairs get detached from this void, contributing to the drag 
force. This behaviour is parallel to the one described in \cite{Josserand} for the case in which a superfluid
modelled by a cubic-quintic equation flows past a rigid obstacle.
It is worth mentioning that the confluence of the liquid which isolates the vortex and antivortex from the void
generated by the moving object qualitatively resembles a splash singularity \cite{splash}, even if the mathematical
details are rather different.

\begin{figure}[htb]
\begin{center}
\includegraphics[width=0.23\textwidth]{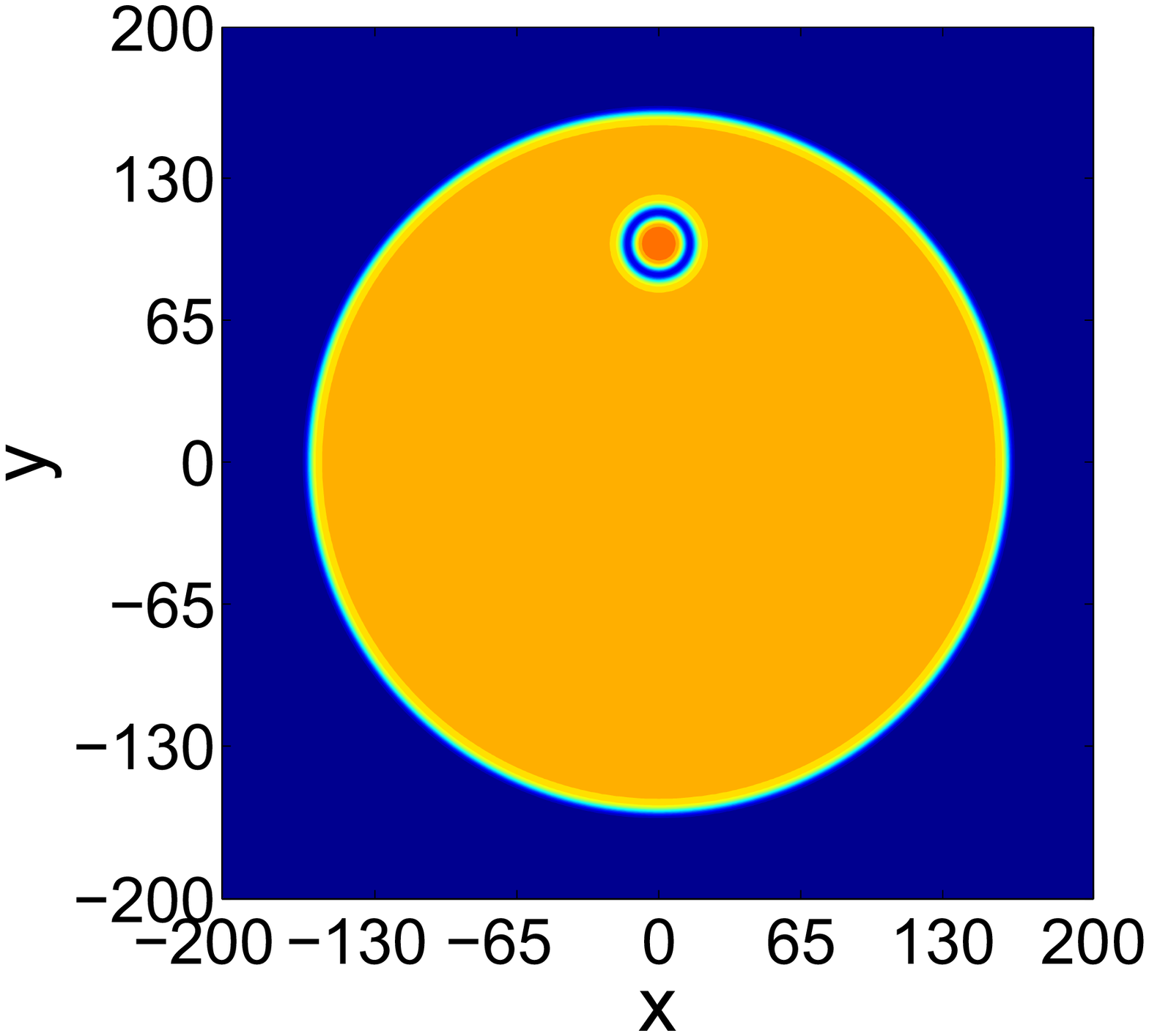}
\includegraphics[width=0.23\textwidth]{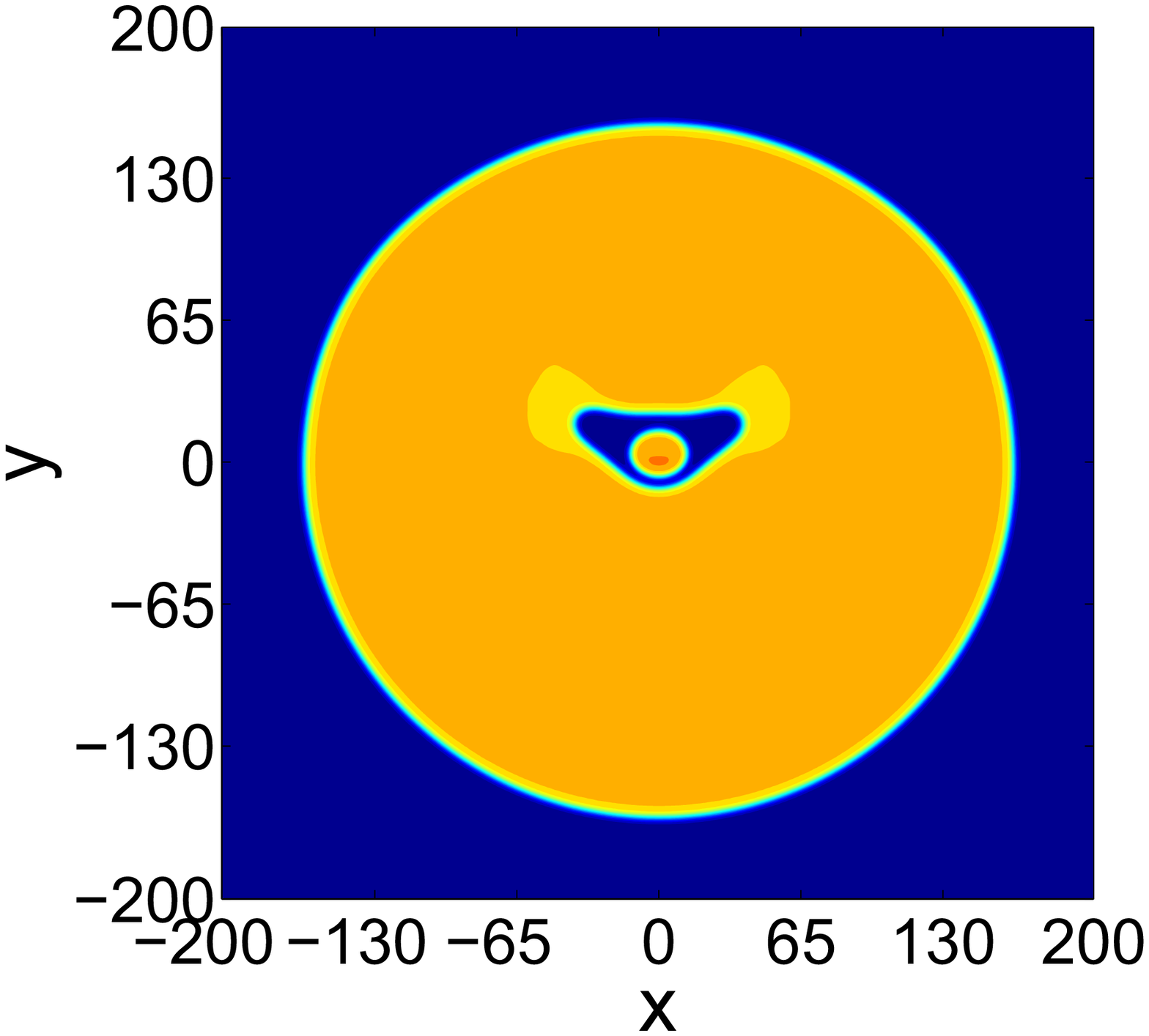}\\
\includegraphics[width=0.23\textwidth]{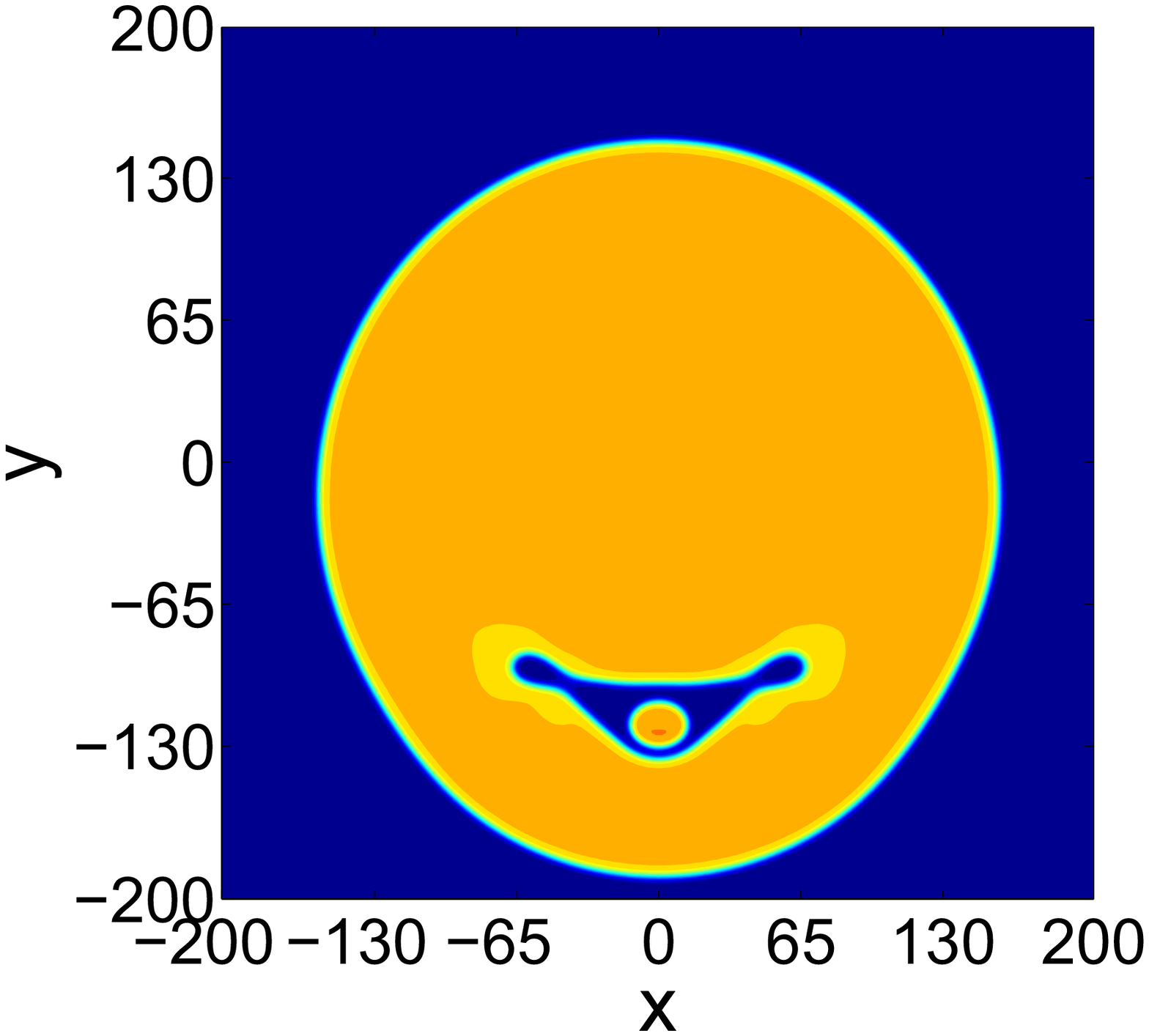}
\includegraphics[width=0.23\textwidth]{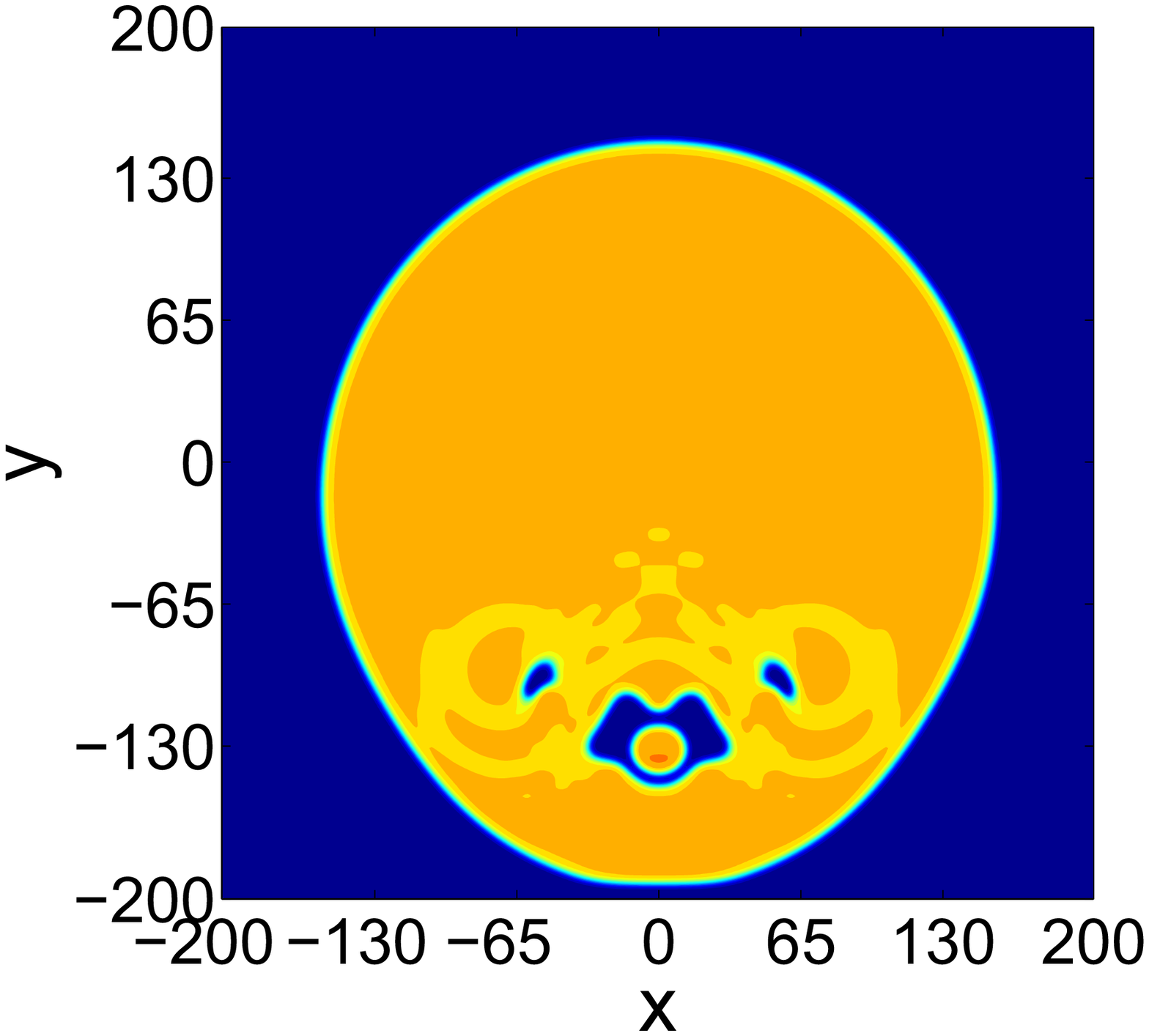}\\
\includegraphics[width=0.23\textwidth]{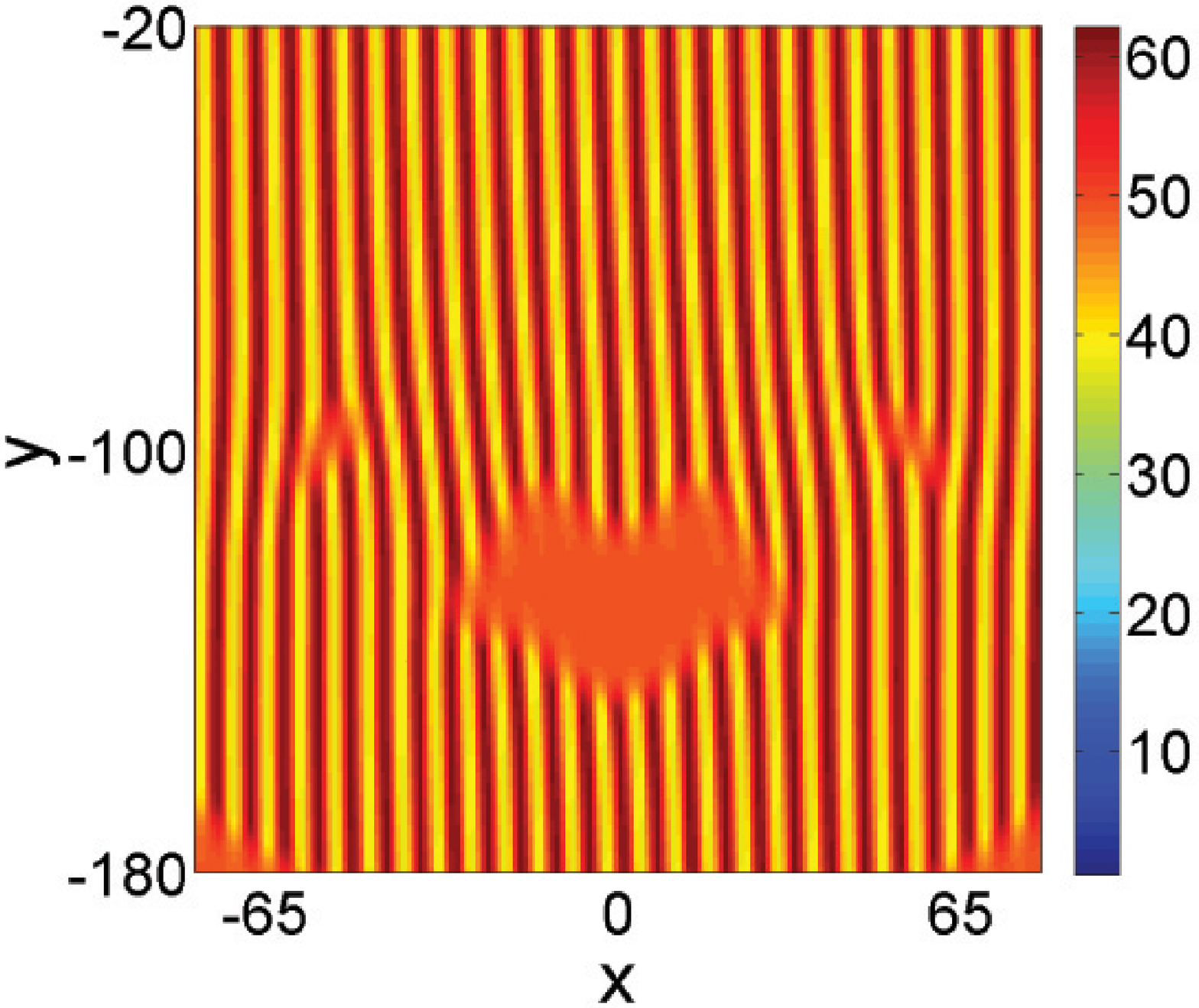}
\end{center}
\caption{Evolution with $g=8\times 10^{-4}$. The initial configuration is
the static solution with
$\beta_1=0.15$.
The four first images  correspond to $z=0$, 800, 1900 and 2000, respectively, and their
color convention is as in Fig. \ref{fig3}.
The terminal velocity is moderate and the moving object leaves a trail
of vortex-antivortex pairs, whose nucleation and detachment contribute to the drag force.
The plot below explicitly shows the phase singularities of the vortex and antivortex, which
appear as fork-like structures in the interference pattern of the wave-function with a plane wave.
Concretely, the image corresponds to $|\psi_2(z=2000) + 7 \exp(1000\,i\,x)|^2$.
}
\label{fig5}
\end{figure}

\begin{figure}[htb]
\begin{center}
\includegraphics[width=0.23\textwidth]{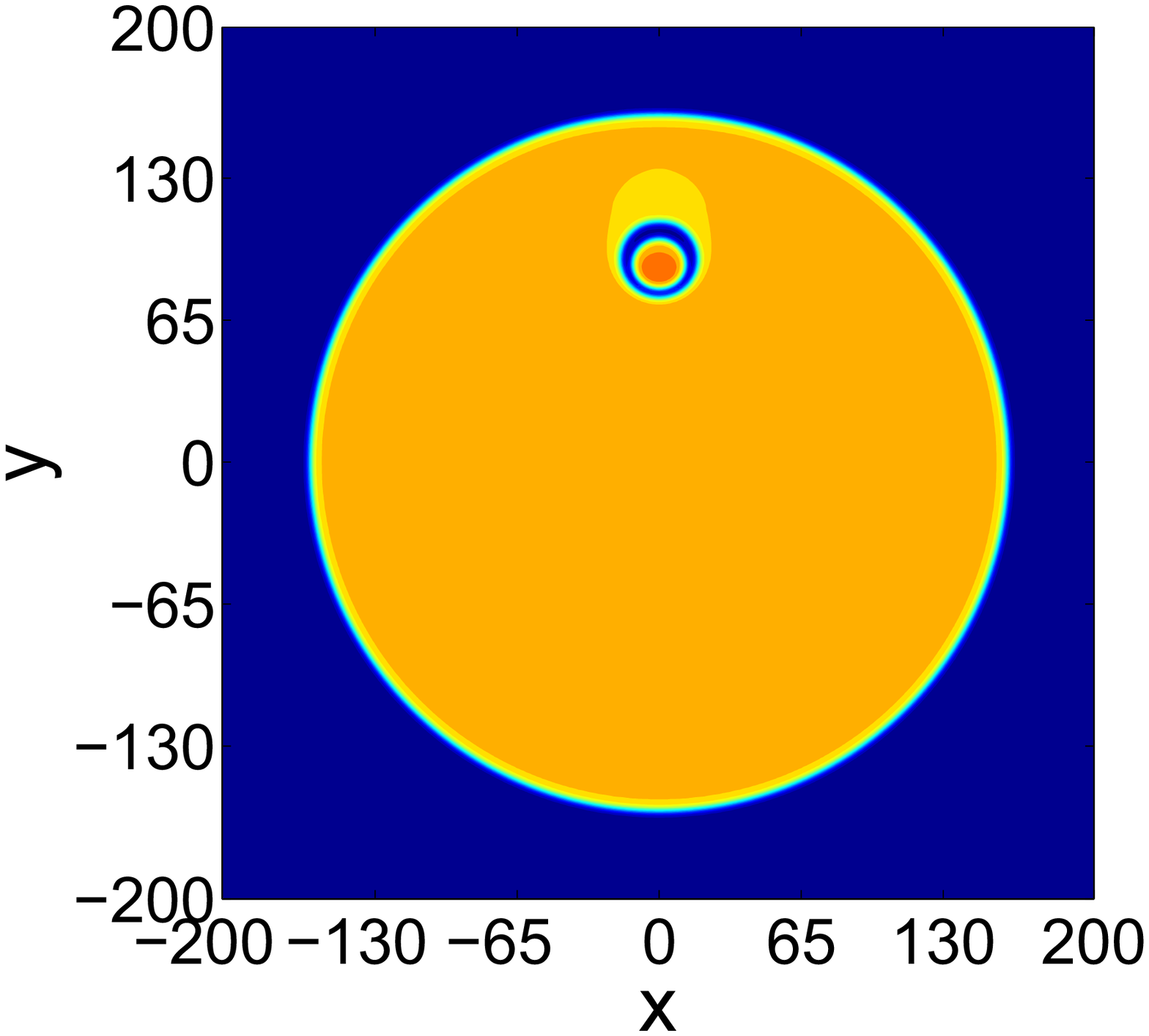}
\includegraphics[width=0.23\textwidth]{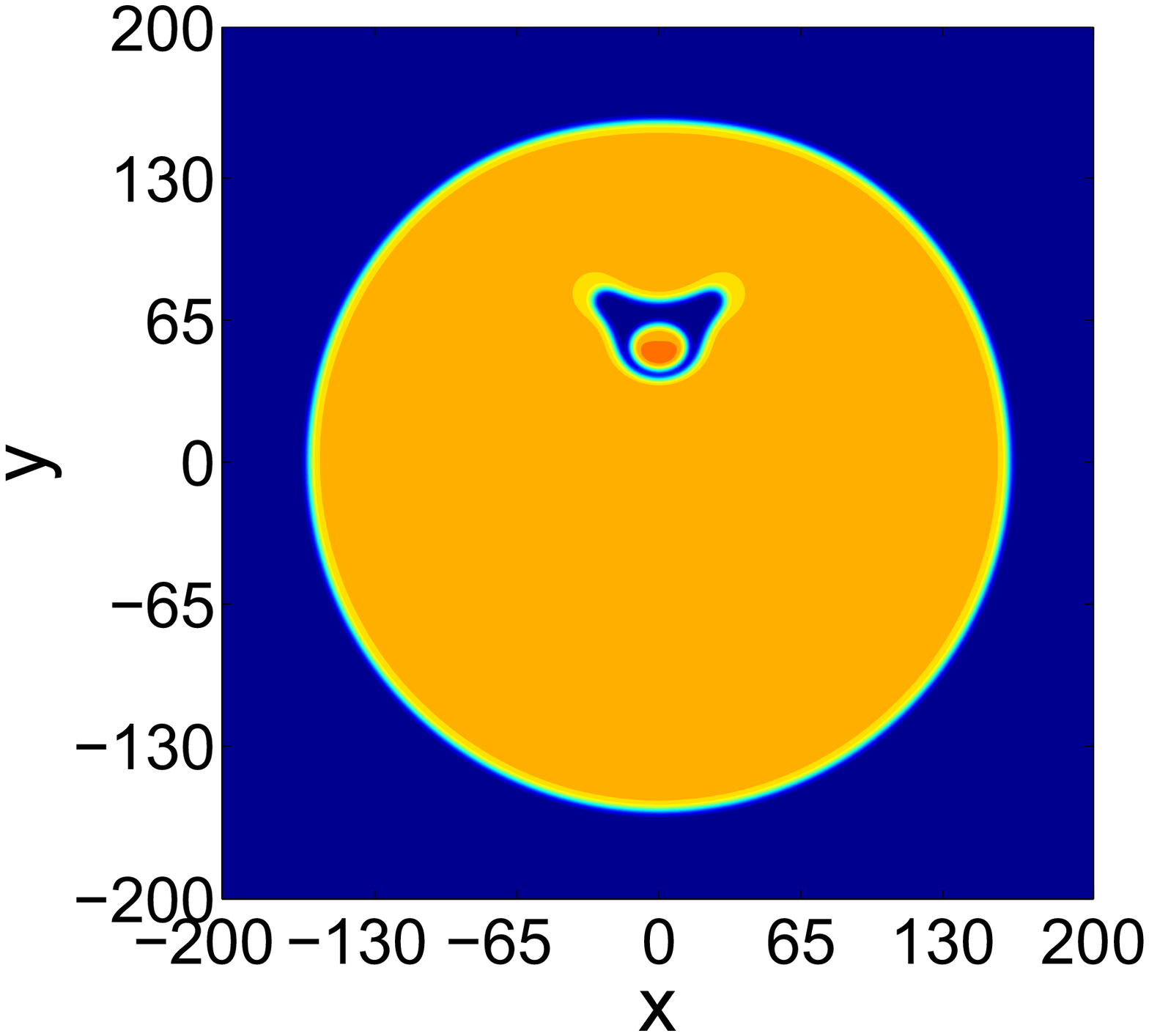}\\
\includegraphics[width=0.23\textwidth]{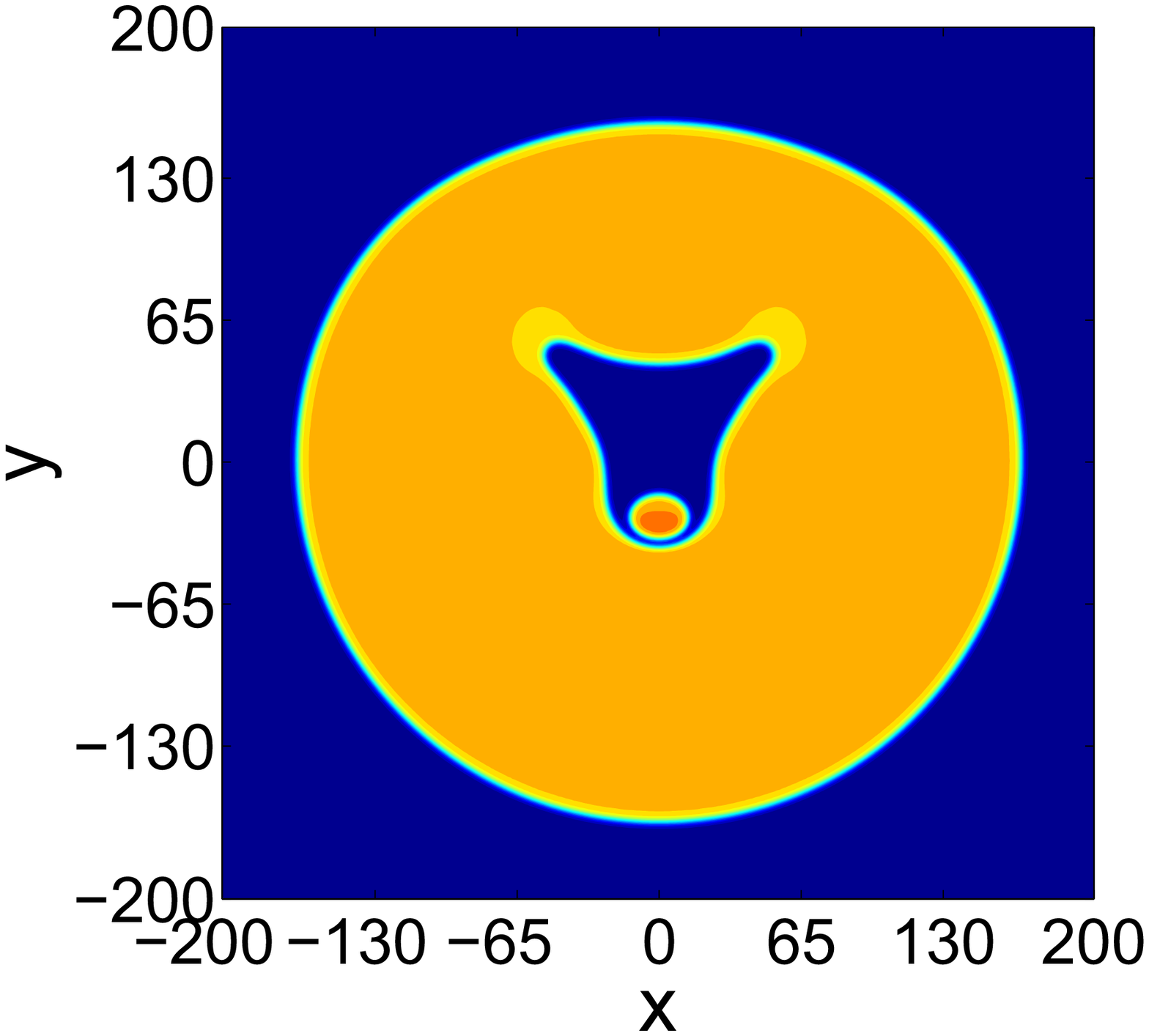}
\includegraphics[width=0.23\textwidth]{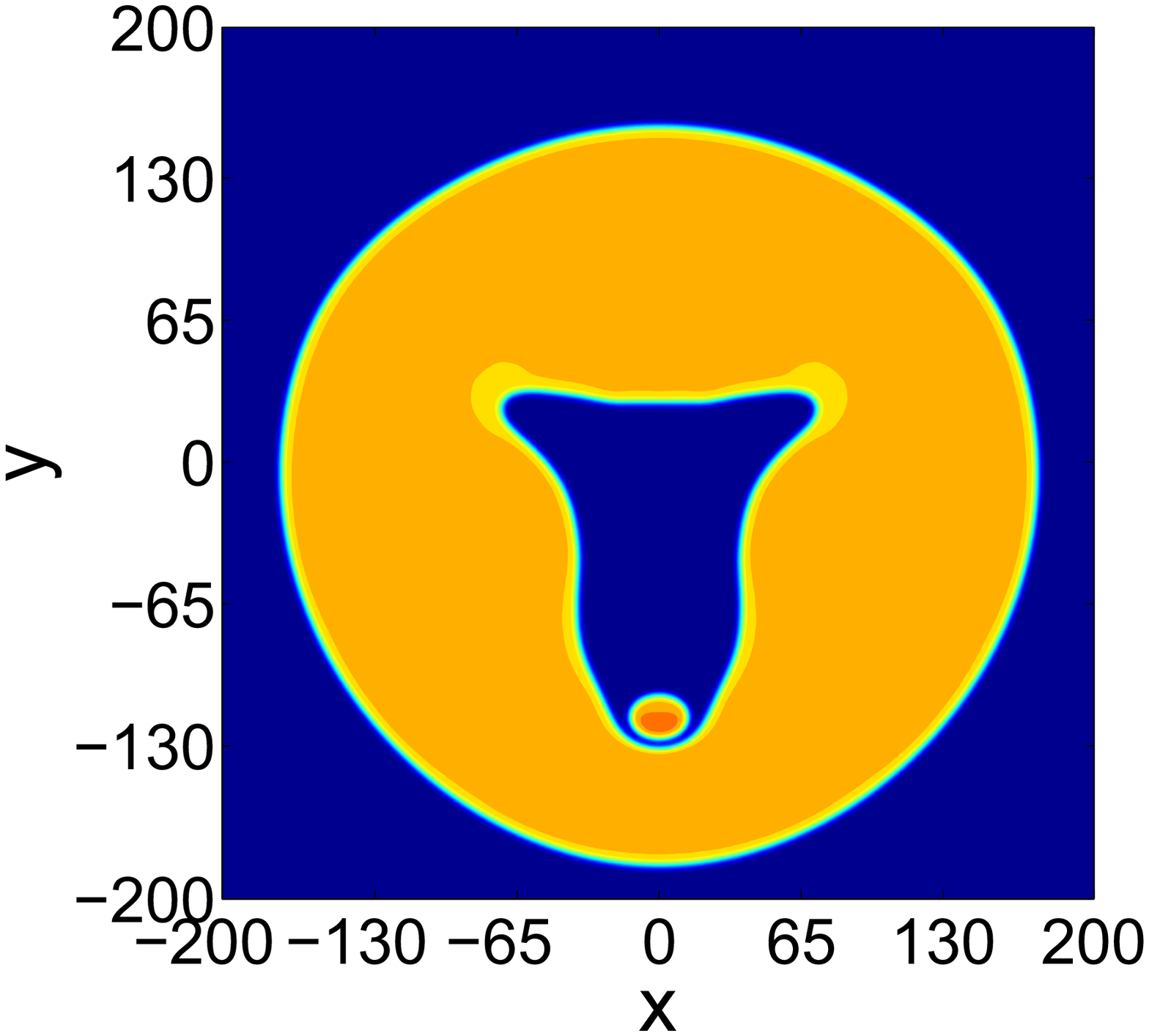}
\end{center}
\caption{Evolution with $g=2\times 10^{-3}$.
The four images  correspond to $z=100$, 300, 600 and 900, respectively.
Colors are as in Fig. \ref{fig3}.
 The terminal velocity is larger than in the previous case and the advance of the
lump of the first species leaves a bubble at its wake.
}
\label{fig6}
\end{figure}

Notice that, as also happened in the set-up of section III, 
the initially round $\psi_1$-distribution gets somewhat deformed. 
As it could be expected, the deformation is greater when larger velocities are reached.
Moreover, for large velocities, the surface tension forces of the surrounding liquid fail
to rapidly occupy the void left at the object's trail and a bubble is generated.
Figures \ref{fig5} and \ref{fig6} show representative examples, see also \cite{supplemental}.

\begin{figure}[htb] 
\begin{center}
\includegraphics[width=\columnwidth]{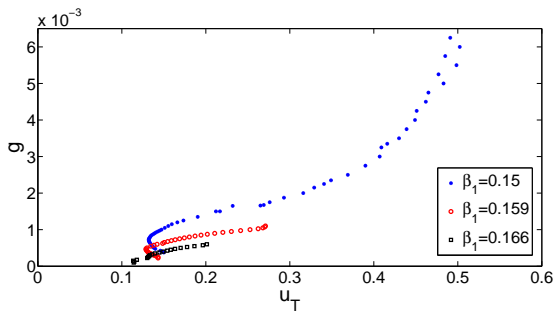}
\end{center}
\caption{Some examples of $g(u_T)$.
}
\label{fig7}
\end{figure}

In this set-up, it is possible to compute numerically the terminal velocity for different values of $g$ and different
initial functions $\psi_1$, corresponding to different values of $\beta_1$ as defined in section II. We  restrict
ourselves to values of $\beta_1$ not far from $\beta_{cr}$ in order to have distributions of $\psi_1$ for
which the analogy to a body within a fluid is applicable to some extent.
We plot some results in figure \ref{fig7}. 

As $g \to 0$ the
value of $u_T$ tends to a positive constant, as it could be expected from the D'Alembert's paradox behaviour.
 This result is reminiscent of
\cite{drag}, even if the set-up is rather different.
 For large values of $g$, the drag becomes quadratic in velocity.
When $\beta_1$ is very near $\beta_{cr}$, the quadratic drag regime already starts at small 
velocities.  Presumably, the reason is that the lump becomes more
malleable in this regime yielding a modification of the qualitative behaviour.

For certain ranges of $g$ and different values of $\beta_1$, 
the results can be approximated by straight lines (notice however that for small $g$, $u_T$ can decrease
with increasing $g$).
This linear growth suggests the possibility of considering a simple modelling of the situation in
which the drag force is just considered linear in velocity. 
A body subject to a constant force and a quadratic drag force satisfies $\frac{d\langle y_1\rangle}{dz}=-g+k\,u$, which
gives $\langle y_1\rangle = -\frac{g}{k^2} (k\,z + e^{-k\,z} -1)$. We have compared the numerically computed
trajectories $\langle y_1(z)\rangle$ to fits of the form
\begin{equation}
\langle y_1(z)\rangle= -a (b\,z + e^{-b\,z} -1)
\label{mech1}
\end{equation}
where $a$ and $b$ are taken as free parameters. It turns out that the simple mechanical model
 is rather precise for the setup of the present section in a large range of parameters. 
\begin{figure}[htb]
\includegraphics[width=\columnwidth]{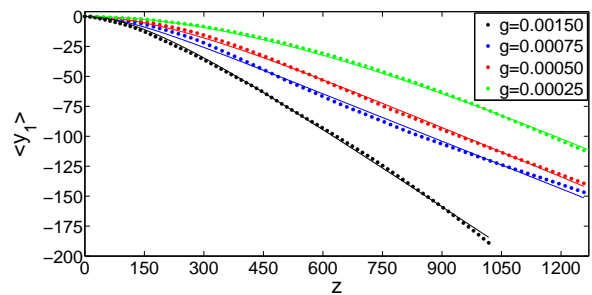}
\caption{Examples of comparison of $\langle y_1 \rangle (z)$ computed numerically
(dots) with fits to the simple model (\ref{mech1}) (solid lines). In all cases the initial conditions
use the $\beta_1=0.15$ solution of section II.
}
\label{fig8}
\end{figure}
This cannot be an exact characterization of the system for several
reasons: first of all, we have seen in section III that there are dragless flows and
(\ref{mech1}) fails to describe them. Indeed, a more complicated dependence of the drag force on $u$ 
was found in Fig \ref{fig7}, which could be introduced in the mechanical model at the cost of losing
simplicity.
 Moreover, since the ``object'' itself
gets deformed while propagating, its interaction with the environment should depend on its form too. 
This explains, for instance, the mild oscillations of the simulated motion in Fig. \ref{fig8} around the solid
lines. As the $\beta_1$ of the initial distribution deviates away from $\beta_{cr}$, the precision of the
simple modelling (\ref{mech1}) declines.

\section{V. Initially separated species}
\begin{figure}[htb]
\includegraphics[width=0.23\textwidth]{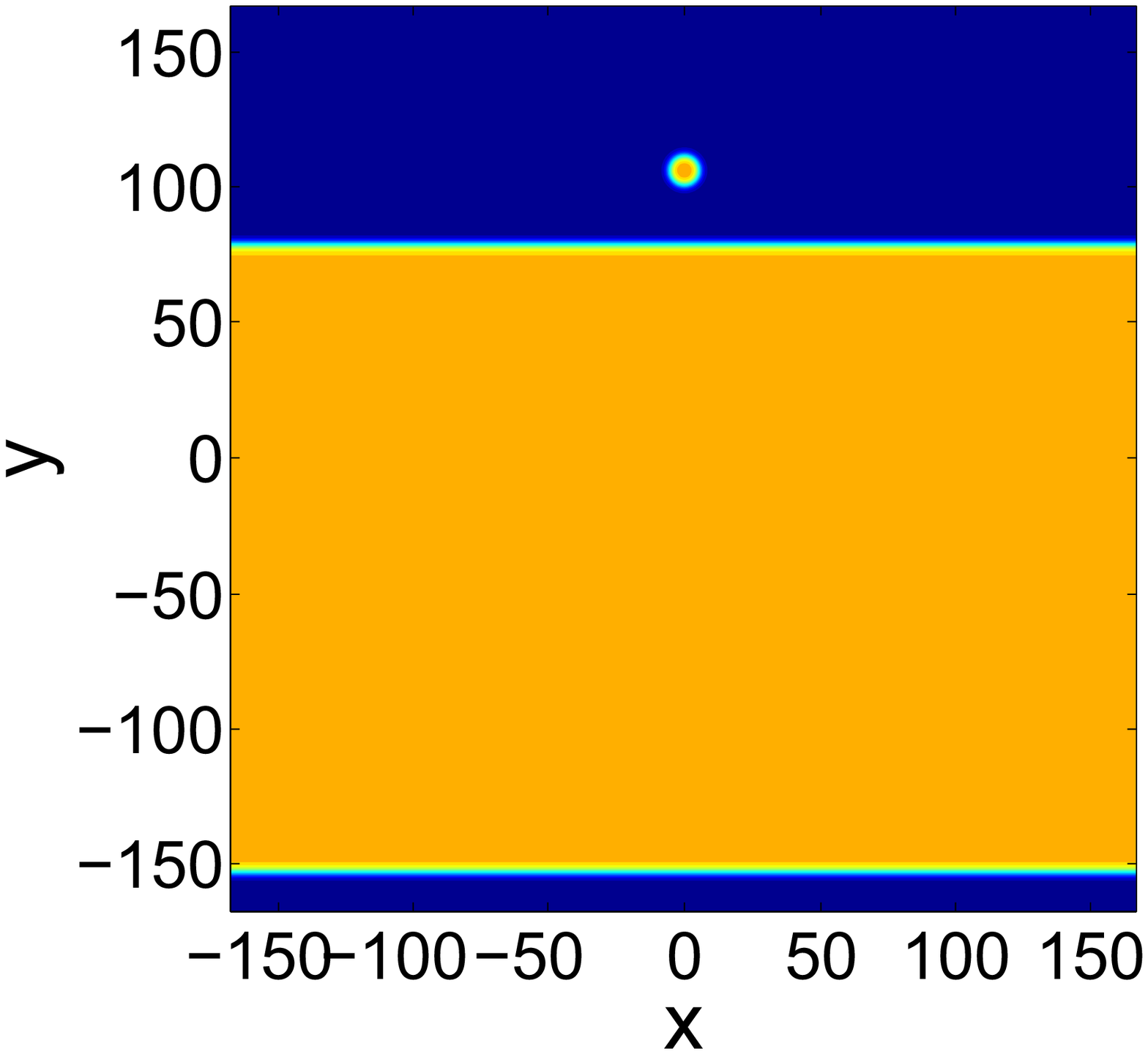}
\includegraphics[width=0.23\textwidth]{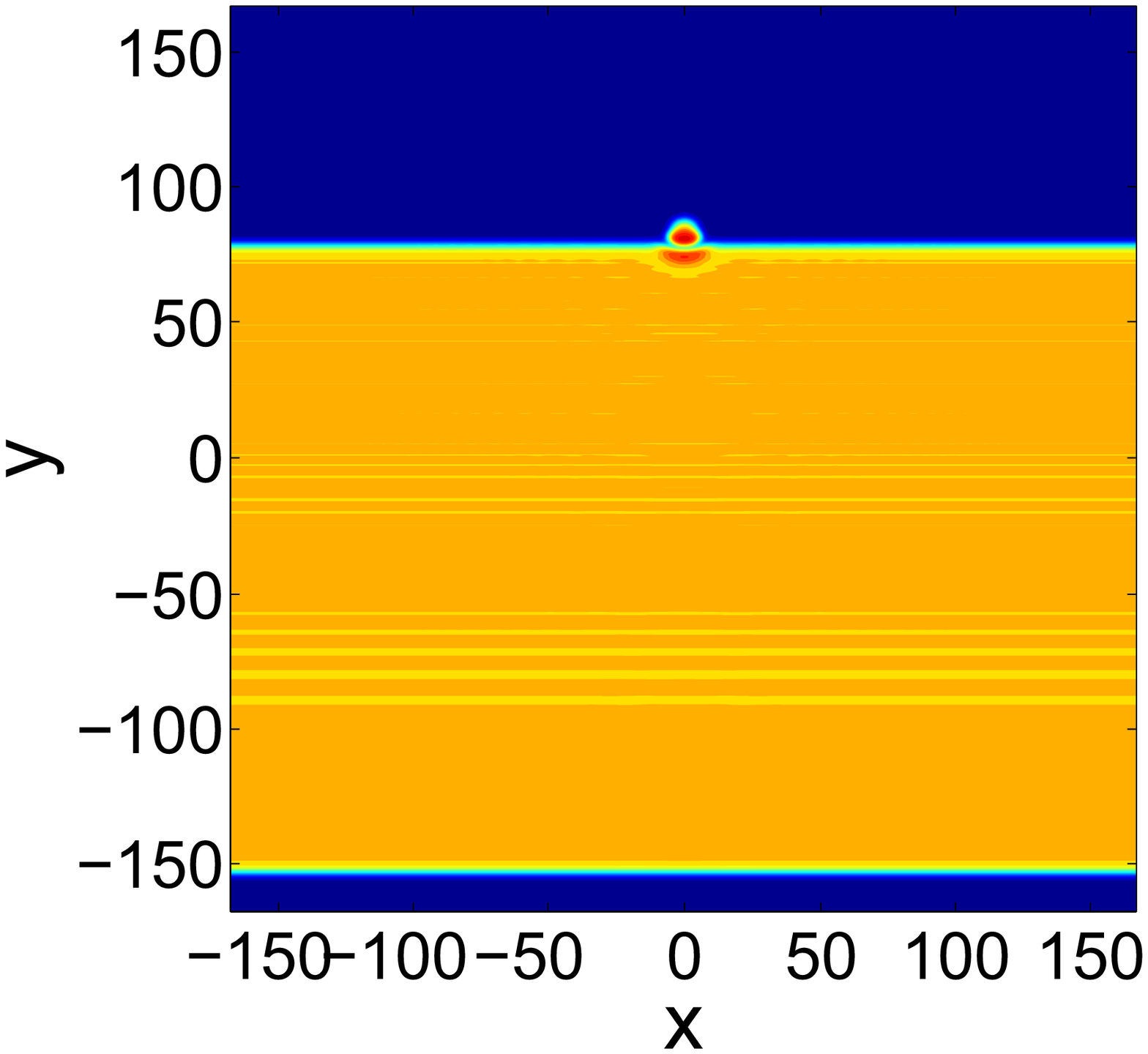}\\
\includegraphics[width=0.23\textwidth]{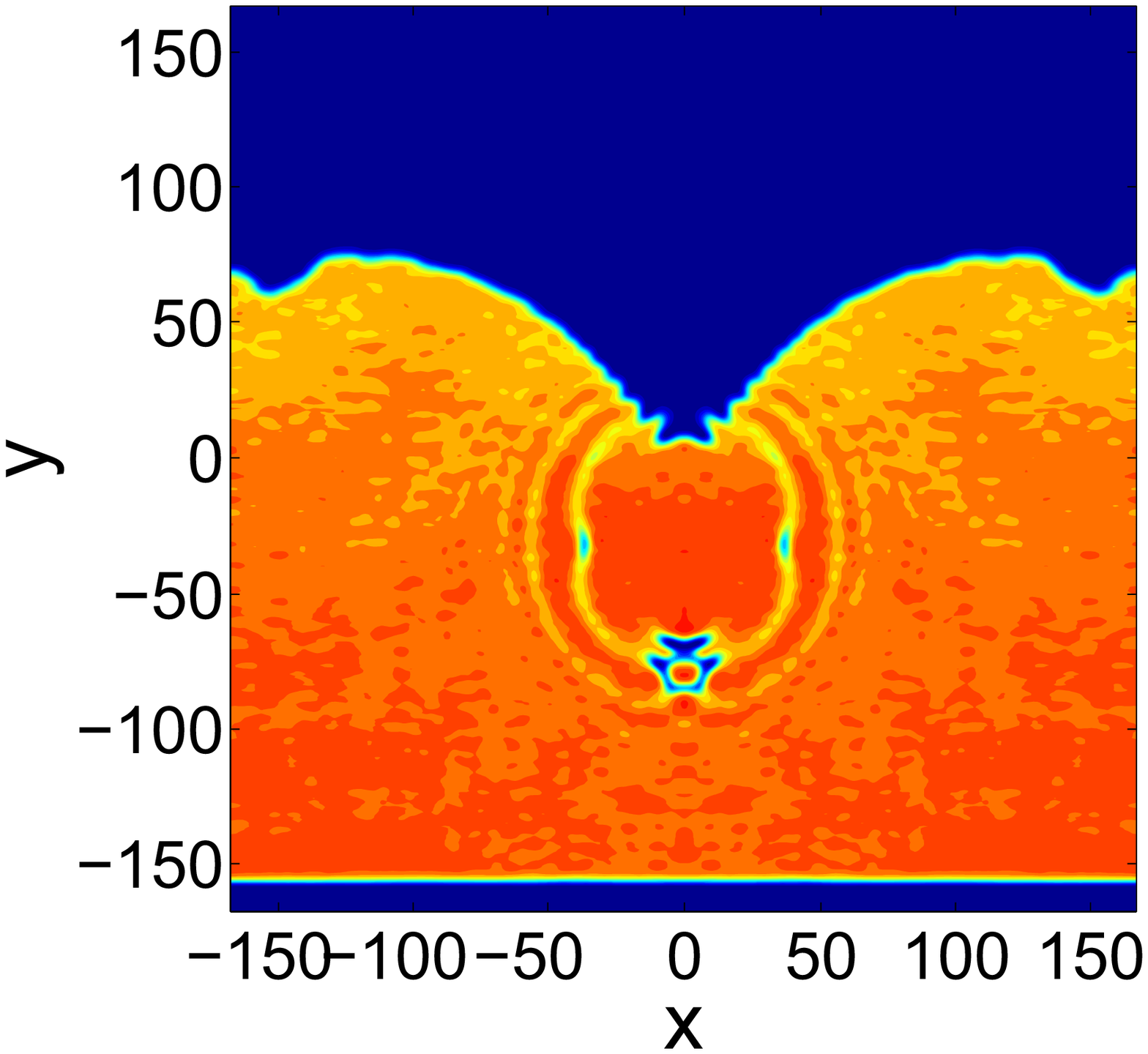}
\includegraphics[width=0.23\textwidth]{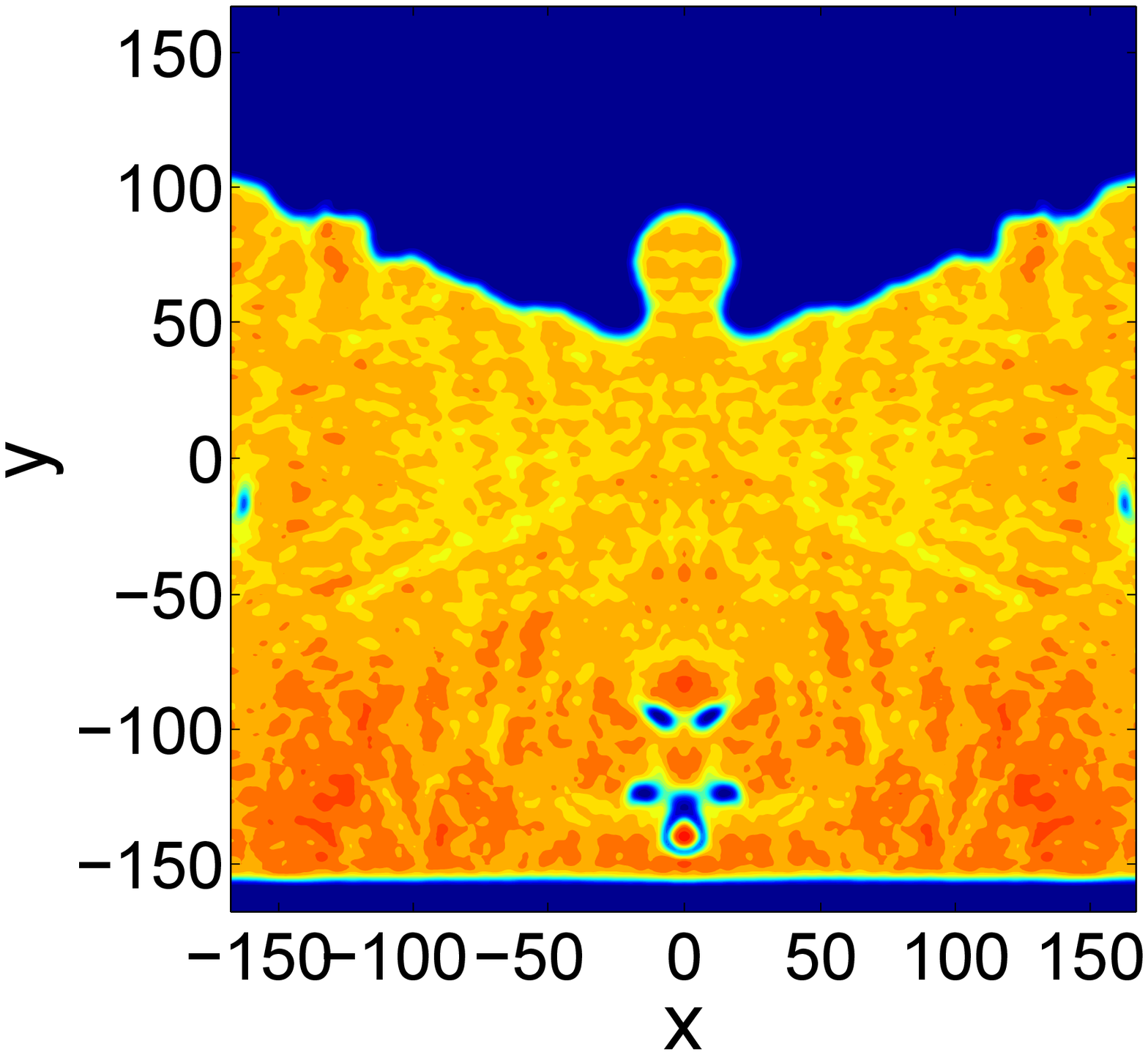}
\caption{A soliton entering the bulk of a fluid, as described by a 
bimodal system of  cubic-quintic non-linear Schr\"odinger
equations. The different images correspond to $z=10,50,500,700$ in the simulation.
Colors are as in Fig. \ref{fig3}.
}
\label{fig9}
\end{figure}

Up to now, we have considered situations in which species 1 is initially within the fluid.
In this section, we illustrate the case in which both species are separated at the outset while
afterwards, the soliton of species 1 enters the bulk of species 2. 
Concretely, we consider equation Eqs. (\ref{eqs}) with a linear potential term $-g\,y\,\psi_i$ for
both species. For species 2, we also include a potential barrier at the bottom. 
Periodic boundary conditions are considered in the $x$-direction.
The evolution
is depicted in Fig. \ref{fig9}, see also \cite{supplemental}. 
We observe that the collision produces surface and body waves. Due to the analogy to a system with
surface tension of the cubic-quintic equations, the behaviour at the surface resembles that of a liquid
hit by an object. Once the soliton enters the liquid, it starts experiencing a drag force as described in
the previous sections. The simulation shows the eventual nucleation of vortex-antivortex pairs related to
this friction process, see the last image in Fig. \ref{fig9}. 
Moreover, since the liquid is somewhat stirred, dark solitary traveling waves
(namely, rarefaction pulses \cite{jr1, rarefaction}) can be excited. Two of them moving horizontally in 
opposite directions can be seen in the plots.

\section{VI. Conclusions}

We have analysed a coupled system of cubic-quintic nonlinear Schr\"odinger equations in order
to understand drag forces in physical systems that can be modelled as fluids within this formalism, such
as the so-called liquid light or certain Bose-Einstein condensates. The two equations correspond to having
two modes, such as transverse polarizations for light or different atomic species. A lump of one of the
species immersed in a larger fluid of the other is subject to macroscopic forces that influence its dynamics.
For small velocities, there are situations in which a D'Alembert's paradox situation exists, namely the
lump moves preserving its form and velocity. Notice however that it would be wrong to say that it is unaffected
by the inviscid fluid, since the energy distribution of the first species does depend on its velocity, as 
shown in Fig. \ref{fig3}. For larger velocities, the drag forces set in. Roughly, it can be said that they
grow linearly with $u$ in a certain region and then grow quadratically
for larger $u$. It is possible to establish an approximate mechanical analogy and consider that the system is just
described by a simple equation for a rigid object subject to a force which only depends on velocity. 
This modelling is accurate to some extent but it is obviously limited since for instance
it disregards
deformations of both the fluid and
the object as described by the CQNLSEs, which also alter the macroscopic dynamics. Overall, these results give
further evidence of the qualitative resemblance of physical systems modelled by the CQNLSE to ideal liquids.


\paragraph{Acknowledgements.-}
We thank D. N\'ovoa for discussions and comments on the manuscript.
The work of AP is supported by the Ram\'on y Cajal program. 
The work of DF is supported by the FPU Ph.D. program.
The work of DF, IO and AP is also supported by Xunta de Galicia through grant EM2013/002.


\end{document}